# Unveiling AI's Threats to Child Protection: Regulatory efforts to Criminalize AI-Generated CSAM and Emerging Children's Rights Violations

Emmanouela Kokolaki[1], Paraskevi Fragopoulou[2]

Foundation for Research and Technology - Hellas (FORTH), Institute of Computer Science
N. Plastira 100, Vassilika Vouton, GR-70013 Heraklion, Crete, Greece

**ABSTRACT**

This paper aims to present new alarming trends in the field of child sexual abuse through imagery, as part of SafeLine's research activities in the field of cybercrime, child sexual abuse material and the protection of children's rights to safe online experiences. It focuses primarily on the phenomenon of AI-generated CSAM, sophisticated ways employed for its production which are discussed in dark web forums and the crucial role that the open-source AI models play in the evolution of this overwhelming phenomenon. The paper's main contribution is a correlation analysis between the hotline's reports and domain names identified in dark web forums, where users' discussions focus on exchanging information specifically related to the generation of AI-CSAM. The objective was to reveal the close connection of clear net and dark web content, which was accomplished through the use of the ATLAS dataset of the Voyager system. Furthermore, through the analysis of a set of posts' content drilled from the above dataset, valuable conclusions on forum members' techniques employed for the production of AI-generated CSAM are also drawn, while users' views on this type of content and routes followed in order to overcome technological barriers set with the aim of preventing malicious purposes are also presented. As the ultimate contribution of this research, an overview of the current legislative developments in all country members of the INHOPE organization and the issues arising in the process of regulating the AI- CSAM is presented, shedding light in the legal challenges regarding the regulation and limitation of the phenomenon.

## 1. Introduction

SafeLine is the only Internet hotline in Greece dedicated to accepting reports of illegal online content. As an official member of INHOPE (International Association of Internet Hotlines) since October 18, 2005, its primary mission is to eliminate CSAM (child sexual abuse material) from the internet and protect children's right to safe online experiences. In 2008, SafeLine joined forces with the Greek Awareness Node and the Greek Helpline, leading to a more unified, organized, and interdisciplinary approach to protecting underage internet users from cybercrime. Through its commitment to tackling child sexual abuse and exploitation, in 2024, SafeLine earned its official recognition as a Trusted Flagger in Greece[3] under the European Union's Digital Services Act (DSA), a status that offered the hotline the mandate to report illegal content more swiftly, contributing thereby to a safer online environment for all.

---

[1] Corresponding author: Emmanouela Kokolaki, Foundation for Research and Technology - Hellas (FORTH), Institute of Computer Science, N. Plastira 100, Vassilika Vouton, GR-70013 Heraklion, Crete, Greece. E-mail address: kokolaem@ics.forth.gr

[2] Author: Paraskevi Fragopoulou, Foundation for Research and Technology - Hellas (FORTH), Institute of Computer Science, N. Plastira 100, Vassilika Vouton, GR-70013 Heraklion, Crete, Greece. E-mail address: fragopou@ics.forth.gr

[3] 2024, October 21, FORTH - SafeLine Hotline: The First Official Trusted Flagger in Greece Under the European Digital Services Act (DSA), https://www.ics.forth.gr/new/15914

An important part of SafeLine's multifaceted action plan has always been the conduct of research on the internet use in schools across Greece, the interpretation and application of legislation related to the protection of children from online sexual abuse and exploitation, together with the identification of legislative gaps, as well as the investigation of alarming trends linked to illegal activities on the dark web. This paper is published as a continuation of the hotline's previous work [1] in an effort to highlight new observations regarding the phenomenon of child sexual abuse through imagery.

Due to the swift evolution of the AI-CSAM and its generation methods, the Internet Watch Foundation (IWF) published two reports reflecting the growing threat of the phenomenon: the first one [2] verifies the presence of over 20.000 AI-generated images on a dark web forum over a period of a month, which depicted more than 3.000 criminal activities. The updated report which was released by IWF in July 2024 [3] appraises what has changed in the landscape since then. In these two reports, the IWF attempted to classify AI-generated CSAM which was detected, in two major categories of illegal material criminalized under the national legislation on the basis of different classification criteria: The Protection of Children Act 1978 (which criminalizes acts related to indecent photographs or pseudo-photographs of a child) and the Coroners and Justice Act 2009 (which criminalizes the possession of "a prohibited image of a child").

At the same, findings from the analysis of the users' discussions relating to emerging threats in imagery were highlighted. Specifically, the research revealed that dark web forums remain the primary location to obtain intelligence on many aspects of AI CSAM, that offenders are able to legally download AI technology and produce AI imagery offline in order to escape detection, while profit-driven illegal activities related to AI-CSAM are already taking place. In parallel, the circulation of pedophile manuals and tutorials on how to utilize text-to-image based generative AI tools was discovered, while the legal status of AI CSAM models remains a complicated matter.

The 2024 report by IWF also underlines the enormous difficulties for law enforcement and private sector to detect and combat CSAM, due to the fact that text-to-image technology is gradually improving, and AI CSAM is turning out to be indistinguishable from real CSAM. This evolution is largely due to the companies' developments in text-to-image creation and the increasing use of fine-tuned models[4] shared in AI CSAM dark web forums. In the course of research of these IWF reports, evidence of open-source AI models misuses was also uncovered, despite the various safety features embedded. What is more, over the course of the 30-day timeframe, IWF scraped all of the accessible metadata on all of the photos shared in this dark web CSAM forum, and assessed them in order to derive valuable conclusions on the models used for the production of these photos. The research proved that a set of over 1000 images had evidence of use of a specific Checkpoint model[5] that was freely available to download, while LoRA[6] (Low-Rank Adaptation) models applied on top of Checkpoint models proved to be CSAM finetuned models. One more study has been published in the field [4], by the International Policing and Public Protection Research Institute of the ARU University, the aim of which was to examine Dark Web forum posts to gain insights on the Dark Web users' views on AI CSAM.

This paper expands our previous work in the area by extending the correlation analysis between SafeLine's reports and domains of illegal sites discussed in the dark web to the examination of whether domain names of URLs reported to SafeLine are identified in dark web forums where users' discussions focus on exchanging information specifically related to the generation of AI-CSAM. At the same time, the analysis of dark web forum posts on the emerging but already highly developed area of AI-generated CSAM enabled us to draw valuable conclusions regarding the behavior of users engaged in the generation of AI-CSAM and emerging trends. By using the ATLAS database of the Voyager system, we managed to analyze a set of posts' content, in order to understand how forum members create and share AI-generated CSAM, what are their views on this type of content, while examining the methods and routes employed by forum members in order to overcome difficulties and impediments encountered during the process of generating such content.

---

[4] Fine-tuning: According to the definition given by the IWF, a machine learning model type in which the weights of a pre-trained model are trained on new data, and therefore adjusted, to perform a secondary task. LoRA models can be fine-tuned – trained on further images. LoRAs/LoRA models can be fine-tuned models of known victims of child sexual abuse, which are trained to produce AI CSAM.
[5] Checkpoint models are large-scale 'base' models that are the bedrock for all image generation. These can be 'foundation' models released officially or they can be models that have been fine-tuned by users.
[6] LoRA (Low-Rank Adaptation) models allow users to use low-rank adaptation technology in order to swiftly fine-tune diffusion models.

As the ultimate contribution of this research, an overview of the current legislative developments in all country members of the INHOPE organization and the issues arising in the process of regulating the CSAM generated by Artificial Intelligence tools is presented, extending our previous work in the area by expanding the analysis of the legislative framework concerning CSAM in all country members of INHOPE [1].

## 2. Generative AI technology: Addressing the Issue

Generative AI is a powerful technology referring to a type of artificial intelligence that is designed to create new content and data, by using models trained on large datasets, to output results that reflect patterns and connections drawn during the training phase [5]. The system's learning process is usually taking place through a combination of human and algorithmic feedback, while AI systems can serve various purposes, depending always on the input data and the machine learning approach employed on each case.

Systems trained on vast quantities of text scraped from the internet, whose purpose is to generate text, fall into the category of Large Language Models (LLMs). They excel at interpreting, processing and generating human language [6]. One of the most renowned LLMs nowadays is for instance the ChatGPT, developed by OpenAI, which has become widespread among internet users [7]. At the same time, there also exist multimodal AI systems, which combine different modalities. One such example constitutes the AI image generators, which include models able to create images from various types of data, such as sound. At this category also belong text-to-image models, which are renowned nowadays and are able to generate images that match the description of the user's textual input. Really famous models of this category are DALL·E, developed by OpenAI [8], Stable Diffusion developed by Stability AI [9] and Midjourney [10], developed by Midjourney, Inc. The models released by these companies are called base models or just models. During 2024, these models released new versions which updated the quality of the output images: Midjourney v5.2, the SDXL version of Stable Diffusion and DALL-E 3 are some of the most incredible advancements made in the field, which were launched at the end of September 2024. These three are all diffusion models, which generate really realistic images.

Generative AI, can also be used for malicious purposes [11]. AI-generated CSAM and deceptive deepfakes promoting misinformation are some of the greatest examples. Specifically, companies with open-source models have the code released to the public, which means that the code is editable, and base models can be trained on further images or be fine-tuned. Despite the benevolent purposes that open-source models may serve, such as transparency and democratic accessibility in AI development, they make it more difficult for companies to implement content moderation policies with the aim of restricting the production of illegal content. The insertion of restricted training data to force the model to generate only the content to which it is exposed, and the embedding of banning prompts, which means that companies restrict the terms which can be inserted by the user to generate an output (through a keyword list for example) are some of the relevant methods that do not always succeed in preventing the generation of illegal material. In other words, the limitations embedded by the companies can be overridden by users who aspire to produce illegal content, such as AI-generated CSAM, since open-source software codes are editable and thus anyone can alter them to suit their needs.

Additionally, base versions of open-source models which contain pornography in their training data, and have no prompt restrictions, allow for pornography generation. Generating AI pornography can also be achieved through websites that are dedicated to providing this service – these often use built-in models. Another route to create AI-generated pornography is found through services designed for 'nudifying' images. A user uploads an image of a clothed individual; the model outputs an interpretation of the individual without clothes. Sites also exist dedicated to providing this service.

## 3. Exploring the dark web with Voyager

As already highlighted by the IWF study in the dark web [2] [3], offenders are using open-source models to generate CSAM, because the access can be obtained offline and on-device, by using freely diverse model versions, while also escaping content moderation and detection. At the same time, perpetrators are using a mounting number of fine-tuned models which are shared in AI CSAM communities, that are also intended for use with the latest foundation models.

Under this study, we dive further into the discussions in the Darknet Forums where processes of AI-CSAM generation are shared and discussed, through the use of the ATLAS database, where a number of Darknet forums are indexed, while no images or videos are harvested. This OSINT-type database contains cached pages, derived through the Voyager system that crawls on an ongoing basis a number of CSAM-sharing forums on the dark web. Extensive analysis is performed on the data for the extraction of entities and relations, that enable the inspection of hotspots without the need for actual dark web access. The text content of the posts, the language used, the names of the forums where the posts originate, the topics discussed in the forums and information about the victims' age discussed, are some of the data that can be explored.

Using the ATLAS database, we initially managed to extract statistical data that presents an overview of the main issues and risks that arise for children, while also highlight concerning trends. Thanks to the custom filters provided by the system, we narrowed the scope of our research to investigate specific issues within the field of AI-generated CSAM, such as models finetuning or the production of realistic depictions. Ages extracted from the posts' text, based on phrases such as "12yo" were detected by the System in multiple languages.

Having applied the filter "AI-generated", we explored forum subcategories specifically referring to AI-generated content and attempted to identify which age groups of minors appear most frequently in this topic's posts. In 2023, the data extracted from the posts' text from the forum subcategories specified above, demonstrated that the age of 10 was the most frequently discussed one (18,14%), followed by the ages of 8 (10,29%), 9 (10,29%) and 12 (10,29%). Detailed representation of the results can be found in Figure 1:. The respective search for 2024 revealed that the age of 6 was the most frequently mentioned one (14,41%), followed by the ages of 9 (12,71%) and 10 (12,71%). In Figure 2:, the exact distribution of the reference percentages of ages is demonstrated. The shift in focus towards younger age groups for AI-CSAM production in 2024 may be linked to the optimization of model usage methods, which, as shown in the analysis below, offer the potential for creating more realistic depictions of illegal sexual acts involving very young minors, in which mainly adult body types engage.

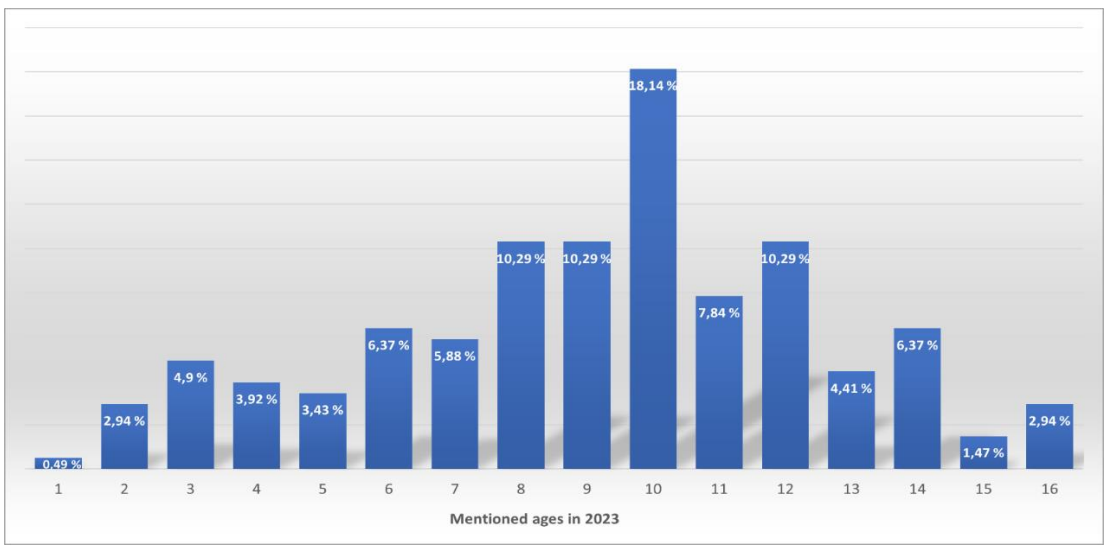

**Figure 1:** Children ages mentioned in dark web forums in 2023.

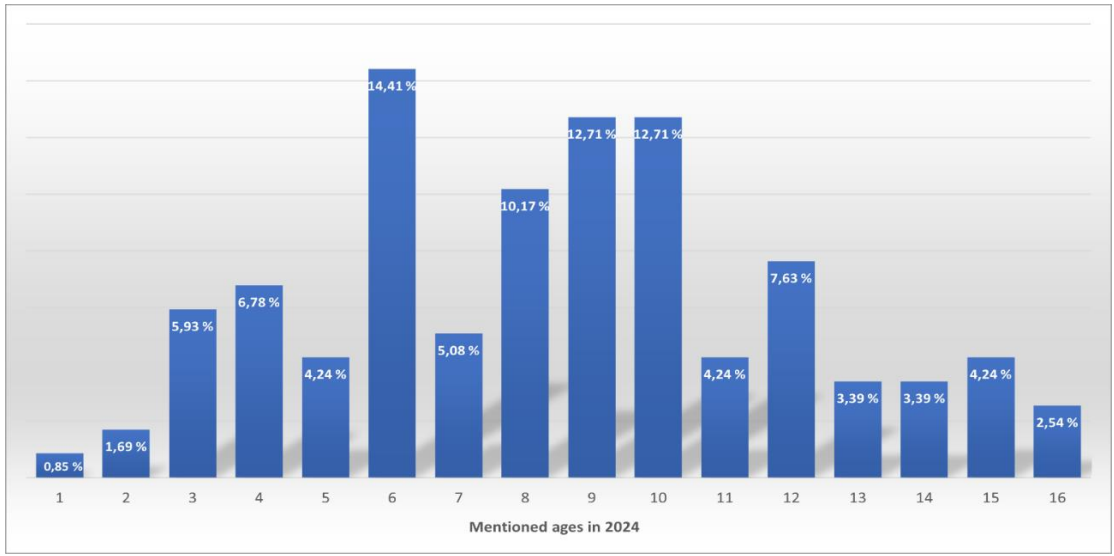

**Figure 2:** Children ages mentioned in dark web forums in 2024.

Thereafter, we attempted to identify existing forums in 2024 where users exchange information regarding the AI topic, and we repeated the process for data from 2023 and 2022. Specifically, by applying the "AI" filter and year 2022, the system indexed 10 forums, where a total of 1.325 posts discussing the AI topic were being published. In 2023, a total of 9.204 posts were found across 10 forums. Finally, in 2024, a significant increase in posts is being observed, with a total of 15.425, while new communities seem to develop, bringing the total to 14 active forums. The corresponding post counts related to the AI topic are presented in Figure 3:, Figure 4: and Figure 5: below. The upward trend of posts related to the AI topic across various forums indexed in the dataset, is depicted in Figure 6:.

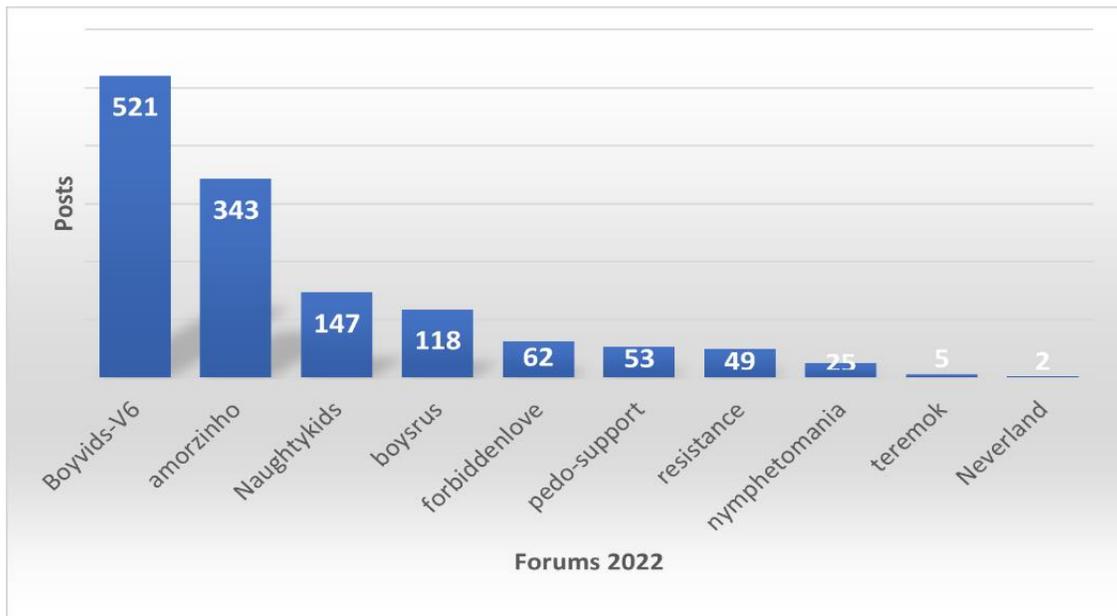

**Figure 3:** Distribution of dark web forums' posts in 2022.

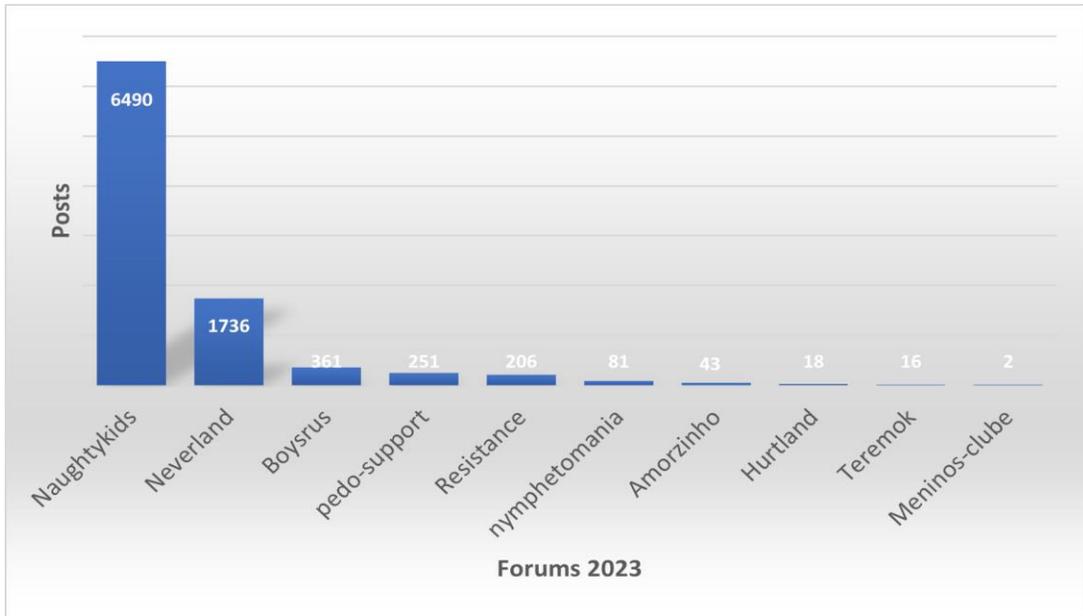

Figure 4: Distribution of dark web forums' posts in 2023.

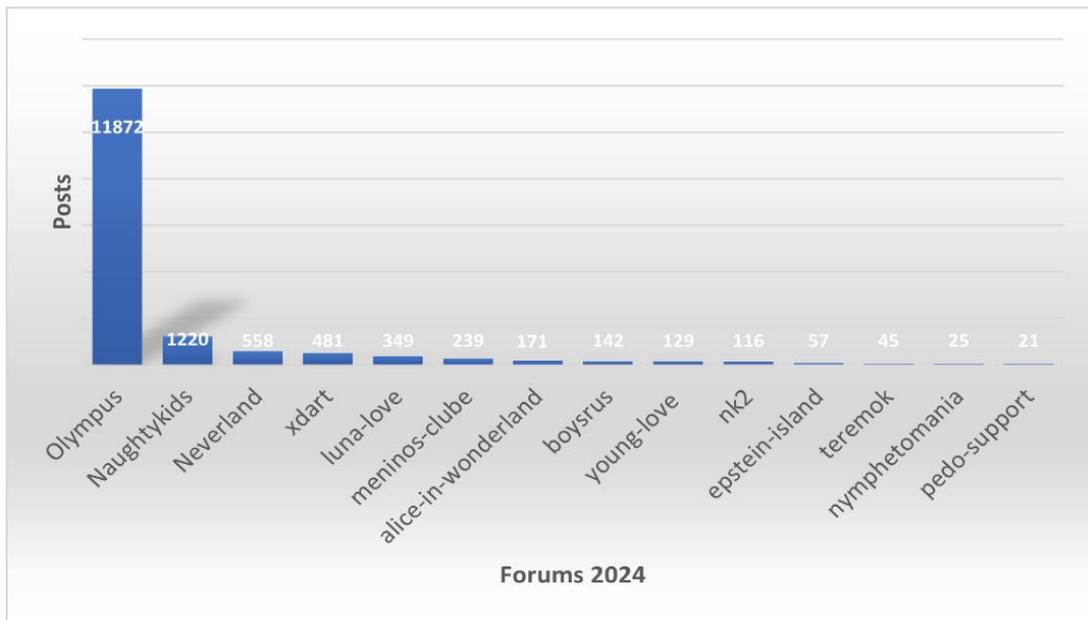

Figure 5: Distribution of dark web forums' posts in 2024.

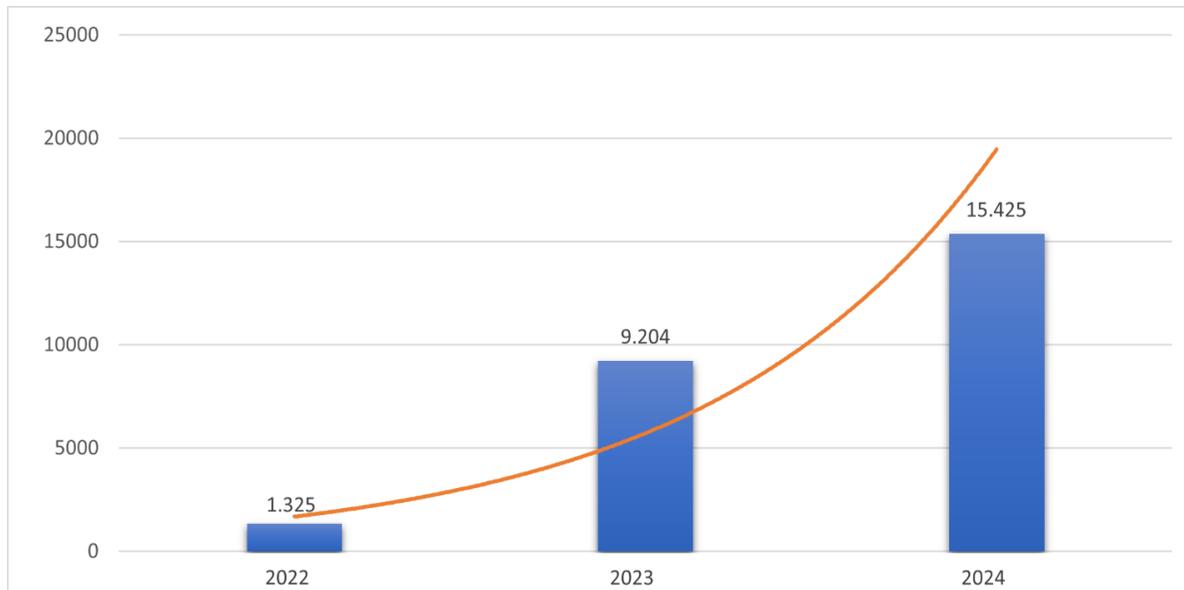

**Figure 6:** Trend of AI related posts in dark web forums.

### 3.1. Correlation Analysis between SafeLine's reports and Dark web data

In order to explore whether there exists a connection of dark web and clear web content, we analyzed 148 verified CSAM reports received by SafeLine in a six-month period, specifically from 01.06.2024 to 30.11.2024, with the aim of revealing whether the domain names of the reported URLs are identified in the CSAM forums indexed by the system described above in 2024, where users' discussions focus on exchanging information related to the generation of AI-CSAM. By applying to our search the filters "AI-generated" and "Forum sections about AI", we reached the sub-category sections we aspired to analyze. Specifically, our research revealed that the domain name (Domain 1) corresponding to 55 illegal URLs reported to SafeLine between August and September 2024 is identified in 32 posts of a specific forum in the dark web, in 2024. Furthermore, another domain name (Domain 2) corresponding to 28 illegal URLs reported to SafeLine during September, 2024, appears in 7 posts of a specific forum in the dark web, in 2024. Last but not least, the domain name (Domain 3) of 7 illegal URLs received as reports by SafeLine during November 2024, is identified in 12 posts of a dark web forum, posted at different times during 2024. It should be noted that all three domain names were identified in posts of the same forum.

In total, three different domain names corresponding to 90 out of 148 reports that we analyzed and were received by SafeLine in a six-month period (01.06.2024 – 30.11.2024) were identified in 51 posts of the same dark net Forum, throughout 2024. Figure 4: presents the number of posts containing the three illegal domains, while in the same graph, one can observe the number of illegal URLs processed as reports by SafeLine, corresponding to these illegal domains.

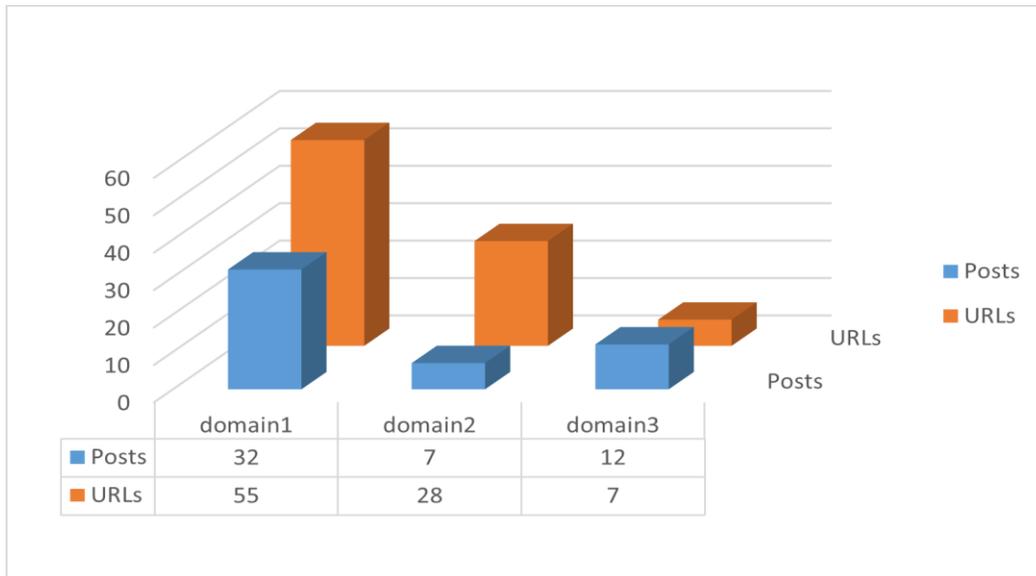

**Figure 7:** Correlation of illegal domains from dark net posts and SafeLine's reports.

Domain 1 was found in a "Preview mirrors" form in the respective posts. "Preview mirrors" displayed in these posts, indicate that by clicking on them one can see iterative steps of the process followed to the output generation, implying that versions of the image can be seen before the final rendering. Relevant archives with the respective passwords were also shared. From our above research, it was proven that some preview mirrors were hosted on a URL of a domain name that was reported to SafeLine.

Domain 2 appeared through links under the title "Preview mirrors", or under threads' titles like "Testing a new LoRA with some HC images". Comments coming from other users indicated that pictures generated depict a person of 10 years old, while others were implying that LoRAs were trained on a specific underage girl that was a model in real life. In other comments a disturbing need of users asking for new images of this girl in "HC situation" (meaning hardcore depictions) was observed.

Ultimately, Domain 3 was included in the list of hosts that tend to go up and down online, the administrators of which however do not seem to easily remove content. This particular domain, in this case, was suggested by a specific user to the community, as part of an exchange of useful advice and tips, to be used with the aim of sharing personal creations and preview mirrors.

### 3.2. Main observations from posts' content analysis

#### 3.2.1 Perpetrators' operations aimed at identifying a suitable model

From a set of 15.352 posts indexed in the dataset in 2024, the content of which was related to the AI topic, a sample of 170 posts was selected to be analyzed. These posts carried a timestamp between May 2024 and November 2024, covering a span of six months. They all included CSE indicators from the text provided by the System in multiple languages ("CSE explicit", "sexually explicit") and were derived by searching into specific forum subcategories, with headings such as "AI-generated" and "AI". Different keywords were also employed in order for specific entities related to AI to be extracted, such as "finetuning", "checkpoint" or "realistic".

In the majority of posts and the respective replies, users were discussing about certain files constituting custom trained models or fine-tuned checkpoints for famous text-to-image generative models. These files and instructions on where to find them were referring to LoRAs and were exchanged between users. For instance, a source where one could find a fine-tuned checkpoint could be online platforms which are dedicated to AI art generation and sharing of custom models and assets for AI tools. The names of the files shared consist of the model's name, the version and the file format for saving the model. They later propose certain advanced checkpoint versions for guaranteed results. It should be noted that the user quotes provided below have been carefully redacted to omit sensitive content and references to AI model versions for security purposes:

User's quote: "*The checkpoint itslelf is a [redacted model's name] mix*", referring to a specific model checkpoint which has been fine-tuned or mixed with styles associated with certain characters.

User's quote: "*This LoRA is based on the [redacted model's information] available on [redacted platform's name]. Use it on that checkpoint, the checkpoint itself is a [redacted information] based mix. If you use it on any other checkpoint expect the likeness to go down*".

Additionally, users share tips and prompt guidance in order for the AI model to strongly follow the text prompt, while certain tags are proposed to act as labels or keywords, in order to help the system generate expected elements based on the user's input. A user gives also suggestions on how to use EXIF (Exchangeable Image File Format)[7] metadata that the prompts contain, since in the context of AI-generated art some tools may embed the text prompt in the EXIF metadata of the output image. The suitable features offered by certain web-based interface for generating images with AI models are also shared to help users extract the detailed information from a PNG file, as seen in the example below:

User's quote: "*The [redacted information] have exif metadata that contains the prompts. Use [redacted information] to extract it*".

At the same time, certain labels, tags and specific positive and negative prompts are shared by some users, in order for deformities to be avoided, while instructions are given on how to open LoRAs and checkpoints and which AI program should be used. Advice on dimensions, on specific combination of words and phrases to guide AI models to generate certain outputs and decrease the age of the depicted character are also provided, while guidance for distinguishing between models that produce realistic results and those that do not, is offered. Moreover, models that generate more realistic outputs of older teenagers, based on the confirmation that they contain datasets of adult genital anatomy are shared.

Additional techniques and settings, such as CFG (Classifier-Free Guidance) that help to steer the generative process of an AI model to improve the quality and the realness of the generated images are proposed. In certain posts, user suggestions to apply a certain prompt in combination with a proposed way to use the CFG appear ("*lower CFG*", "*higher CFG*"), with the aim of achieving varying results, from realistic to more creative ones. Subsequently, they showcase the results of their experimentation by sharing sample images with the community. Along with this, it seems that this community's users employ highly sophisticated methods to exploit the models' capabilities, since they exchange information on the model's configuration files or parameters to control the way a model runs during the training and the inference stage. Examples of relevant user comments are provided below:

Users' quotes: "*Negative prompt: bad anatomy, vaginal insertion, open mouth*", "*Also, please post your creations in the comments, that always motivating for me and others*", "*What's your [redacted model's name] config bro? These pics look sooo real*", "*I was able to track down the 2 loras I was missing*", "*Is there any chance you could share any training details? I am very interested in the base model used, number of pictures in the dataset and your captioning method. Also, did you use any special training parameters in ai-toolkit*", "*Would you post your captioned dataset? Then we could try diff training settings and see what works the best*", "*Advice on [...] says little negative needed but I started with nothing and ended up with many lines of things just to keep adults and deformities out of the images. Some models need long strings of child related words to balance the influence of loras but results are worth it*".

Admiration for the content created by some users is also openly expressed: "*seems like you have managed to eliminate alien anatomies, the [redacted sensitive information] are properly constructed…*", "*Damn, your model outputs look fantastic! It would be real treasure for inpainting. Underage [redacted sensitive information] and especially [redacted*

---

[7] EXIF metadata refers to information embedded in image files, which provide details about the image and the conditions under which it was taken (such as location data, device information etc.).

*sensitive information] are very hard to do with current models*". Additionally, there is a need to bring fantasies to life that relate to real children actors of films of the past: "*What former child stars do you want to see?*". They also share passwords to decrypt the links and extract files by decompressing, while they also share versions of different variations: "*No pubic hair version: link*", "*With pubic hair version: link*". Considering the context of the posts, this community's users seem that use advanced techniques to leverage the AI models' potential and tailor the capabilities they offer to the production of customized AI-CSAM: "*...you can force models by using [redacted words pertaining to illegal acts related to children]*", "*When using words in the prompt like lusty then I need to add the words [redacted words characterizing children] - to drag the age down*", "*I can control the object, the size, the color of it as well as the ethnicity and age ...those are just test to see how well my dataset was applying during training.*".

Some users share techniques in order to reach a more realistic output by proposing a two-phase process for improving AI-generated CSAM: at first, they suggest a clean and preliminary version of the output, as the phase of the data preparation, which corresponds to the generation of a structured result clear of any irrelevant details, pertaining to the basic model output. As the next step it is proposed to use a more advanced model to refine the output, making it more realistic and accurate. Other users change the sequence of using different models and rearrange it in order to conduct tests that will yield the best possible result. Details on how to merge different companies' models are also shared. Examples are provided below:

User's quotes: "*my [redacted Model's version] is NOT meant for realism...it excels at making bodies, young [redacted sensitive information], etc. So first clean version with [redacted AI model's name] then you can re-inpaint with the [redacted AI model's name]in order to bring out a more realistic result*", "*I tried this other [redacted AI model's name] realistic merge, between an [redacted] model and a [redacted] model I guess?*", "*A bit of [redacted model's name] at the end, just to get a bit more realistic face*", "*Maybe merge of [redacted model's name] and [redacted model's name] would be a good idea*".

During our research, we also came across a post where a user was mentioning "*in my opinion [redacted model's name] with same prompt and seed really gives the worst quality...but...gives the best quality of skin textures and image generally*". In this post, a combination of a prompt and "seed" is proposed in order to achieve the best AI output: the prompt instructs the model to produce a specific result, whereas together with a specific seed proposed, which relates to how AI handles randomness in its generation, the content and the variability of the AI's output can be defined. Another piece of advice regarding the same issue comes from another user who contributes to the original post with their own response: "*No inpaint, just the prompt and no retry with seeds*".

Archives, respective pass codes and preview example files are also shared, which contain images produced with certain AI models: "*This batch of 258 AI generated fake pictures is with lots of little girl faces [redacted sensitive information]*", "*Without LoRA, the differences in anatomy between the checkpoints are more noticeable*" while this user afterwards shares "*samples with raw checkpoints without LoRA*".

Another common trend identified in these dark web forums is the use of Large Language Models to refine the prompts. For instance, a user contributes to the community the information that: "*Using one of the new language assistance models for creating prompts might be a solution. Most work has been a lot of OT picture sets and finetuning, so it leans more into hebe/pedo then late teen/adult without any prompting, see example below. As you can see it delivers out of the box and also that no embeddings, loras or styles are used*". In such contexts "OT" signifies the original material, implying that real material has been used in order to train these models. Further along, in the same thread, another user's post appears, in which they mention that "*i am working on adding a large language model (LLM) to the workflow. Basically, trying to reverse-engineer what [redacted model's name] and [redacted model's name] do. The LLM creates complex detailed prompts that can assist the [redacted model's name] models to create more realistic pics and settings*".

**3.2.2 Perpetrators' attempts to bypass technological safeguards designed to prevent misuse**

A key issue that concerns the users of dark web forums where AI generated CSAM material is circulated, is finding ways to bypass technological barriers. They usually talk about platforms which can be installed locally, that do not rely on cloud-based services, and allow users to run AI models. At the same time, users share tips on the model's restrictions put in place to prevent the software's misuse, ways to bypass them and routes to avoid activity detection. They are also discussing about the specific attributes some models have, such as parameters that are quantifying the scale of the model during the training phase and determine the complexity and efficacy of a model in capturing more nuanced patterns in data. Relevant examples identified during the research are presented below: "*I installed a local LLM and am running [redacted model's information] and having fun with it. I could not jailbreak the specific ones that were created for improving [redacted model's information] prompts but I did get an 8B LLM one to come pretty close*", "*Always close internet when extracting files or viewing videos*", "*There's an example of how to break safeguards*", "*The [redacted model's information] model is trash and heavily censored. I think we have better hope with [redacted AI style] or [redacted AI model's name]*", "*[redacted model's version information] isn't going to take off honestly. Mostly because of the incredibly restrictive licensing that will make finetuning basically worthless for the people who do such things. [Redacted model's name] is reaching good levels of maturity due to how long it's been out for.*".

Also, specific algorithms are shared, while users are identified as persistently searching for content examples depicting specific sexually abusive acts against children in order to be able to get a reference. A disturbing need to generate images where a toddler's body could be adapted and tuned to appear as if it is realistically engaging in sexual acts that the body of adults could actually perform, is observed. The content of the conversations is even more grim when we come across threads with titles such as "*Testing a new LoRA with some HC images*", where "HC" pertains to hard core images. Some users explicitly describe their fantasy: "*I love [redacted phrase] girls (10-13) years*".

Having added the "Fine-tuning" tag in the filtering process of the Voyager system to narrow even more the posts' topic, we discovered that the process of models' refinement and optimization is connected to conversations for bypassing safety measures. Regarding the obstacles that arise due to the safety features, a user advises that "*at worst we can always create with [redacted model's name] and finish processing in another product*". In parallel, another user states that "*it's not possible they cleaned the dataset to remove words that cross-reference anatomy relating to CP, however no database of that size is 100% cleaned before modelling begins*", while adding later that "*have just shotgunned anything later on that might be used for CP for safety, because they will exclude valid data of non-CP to get all the CP*". From this post's context, it is clear that the user is referring to the implementation of a practice, the purpose of which is the identification of content related to CSAM, even if it results in the loss of valid, non-harmful data in the process.

The desperate need of these users to overcome the technological barriers posed by AI models in order to achieve their goal, is clearly evident from their relentless pursuit of LoRAs, since the base models do not draw a good quality of details: "*The genitals are not very good yet, I need a good lora…*", "*It seems to me that this kind of sensitive data like realistic anatomy of child/adolescent body and genitalia will never appear in publicly available checkpoints. So definitely a good LoRA will always be needed for our purposes*". They also seem to be disappointed by the clear net instructions and sources to handle AI models for the generation of AI-CSAM. Consequently, it is observed that tutorials and pedophile guides are being shared within these darknet forums: "*So far my search for answers on the clearnet has not resulted in anything that seems to help me*", "*Checkout the tutorial workflow thread here: [redacted information]*", "*Here is the entirely of the generation info*". Tutorials usually include positive and negative prompts, proposed image size, model, CFG (Classifier-Free Guidance scale)[8], type of Variational Autoencoders (VAEs), embedded in the model's architecture or used as external component, proposed interface or API to run the AI model locally, supplementary use of a LoRA or not.

---

[8] CFG (Classifier-free guidance scale) is a parameter that controls the degree to which the image generation process follows the given text prompt.

## 4. AI-generated CSAM legislation – Countries' Overview

### 4.1. The problem

According to the definitions of the INHOPE's Universal Classification Schema (see below in Table 1: ), AI-generated media depicting CSAM, otherwise known as synthetic media or colloquially known as "deepfakes", belong to the category of realistic media which contain either a real human being or a person that is indistinguishable from a real human being to the observer. As demonstrated by the above research on the dark web, it is evident that AI-generated child sexual abuse material contributes to the normalization of offending behavior, potentially leading to a more permissive environment for perpetrators, greatly increasing the risk for children. In addition, AI-generated CSAM amplifies the potential for the re-victimization of known child sexual abuse victims, given that AI applications are often trained on the basis of previously known CSAM or CSEM (Child Sexual Exploitation Material), facilitating offenders to create customized AI-generated CSAM [12]. At the same time, accessing CSAM frequently constitutes an initial step towards actual abuse, regardless of whether the media depicts an actual abuse or not.

The rapid technological advancements in conjunction with the fragmented legislative approach in the field of criminal law complicates the adoption of a coordinated response to the phenomenon. However, some States are moving quickly to regulate the AI field and revise their legislations on CSAM crimes in order to ensure that AI-generated CSAM is criminalized under their jurisdiction.

Due to the fact that the majority of the legislative frameworks globally have not yet been updated in order to explicitly criminalize AI-generated CSAM, the analysis below shall also include provisions criminalizing simulated or virtual CSAM which is a term that encompasses all forms of material representing children involved in sexual activities, with the particularity though that the production of this material does not involve actual abuse of real children, but is artificially created in order to appear as if real abuse with real children is depicted [13]. Although the term "child pornography" should be avoided because it fails to describe the seriousness of the abuse from the child's perspective, it is still used in the legislative frameworks of some countries [14]. The inclusion of this term though in the analysis below is selected in cases where a reference to specific legislative provisions explicitly including the term is made.

As already thoroughly explained in Section 3 of this paper, AI image generators are trained on huge volumes of existing images, usually scraped from the web, allowing images to be created from simple text prompts. They can also generate videos where a person is depicted doing things they never did or swap the person's face onto a video depicting another individual. Criminals who use this kind of technology in order to create AI-generated CSAM proceed also to fine-tuning of older versions of AI models to create illegal material of children. This process would involve feeding a model with existing images containing some form of sexual abuse or images of people's faces, which then allows AI to create images of specific individuals [15]. The outcome of this process is that most of the AI-generated CSAM are realistic and indistinguishable from real CSAM.

The problem is not merely a question of explicitly criminalizing the AI-generated CSAM in every national legislation. The furious spread of this phenomenon entails also new legislative challenges, since traditional understanding of criminal responsibility appears to be insufficient to address content provided by AI systems, the production of which do not presuppose direct human involvement. Specifically, traditional legislation often relies on human intent and action in order for a crime to be committed, meaning that the revision of the respective laws is indispensable in order to encompass acts carried out by AI algorithms. The question of whether the programmer, the owner of the AI system or the user of it should bear legal responsibility becomes extremely crucial within this context.

The concept of intent of committing a crime is an elemental component of serious offences in national and international law [16], despite the fact that the concept of intent is given a different meaning under various legislations. At the same time, not all the penal codes include a specific definition of the concept of intent, which means that academics are left to analyze the notion. Among the different forms of intent that may exist under each jurisprudence, it is worth mentioning the classic definition of general intent under the French law: "Intent, in its legal sense, is the desire to commit a crime as defined by the law; it is the accused's awareness that he is breaking a law [17]".

The study of criminal intent is caused by a relevant need for correctly classifying crimes, determining the criminal's degree of guilt, and imposing the adequate penalty corresponding the offender's guilt [18]. One of the fundamental problems when it comes to harmful actions or illegal content produced by an AI system is the attribution of criminal liability, given the fact that AI itself has not "intent", is often self-learning and sometimes evolves in unpredictable ways. This is exactly the case of tools for text prompt-based image generation which were found to have known CSEM in their training data [12], facilitating offenders to create AI-generated exploitation material [19].

In Table 1: , a section from the INHOPE Universal Classification Schema is presented, where core crime characteristics have been turned into definitions aligned with any legislative criteria worldwide.

Table 1: Section with Definitions from the INHOPE's Universal Classification Schema[9]

| | |
|---|---|
| **Realistic** | Media contains **a real human being,** or **the person depicted is indistinguishable from a real human being to the observer**. <br> This would also include real-world images even if the face of the person is hidden in whole, or in part, through cropping, obfuscation, sanitation, or media created with the assistance of artificial intelligence technologies, commonly referred to as **AI-Generated media**. |
| **Manipulated / Unrealistic** | Media is obviously not an accurate representation of a real person or event. These media files contain elements or portions of people, which to the observer are indistinguishable from real human beings, but are easily identifiable as having been digitally created, altered or changed. |
| **Synthetic** | **Synthetic media** (also known as **AI-generated media**, media produced by generative AI, personalized media, personalized content, and colloquially as **deepfakes**) is a catch-all term for the artificial production, manipulation, and modification of data and media by automated means, especially through the use of artificial intelligence algorithms, such as for the purpose of misleading people or changing an original meaning. |
| **Computer Generated Imagery (CGI)** | The use of computer graphics to create or contribute to images in still art or video media. |

In the remaining of this paper, beyond the individual legislative analysis for each country presented below, under Table 1: one can see a summary of the legislative developments in all INHOPE member countries regarding the adoption of AI-specific laws or policies, as well as the potential existence of specific criminal provisions for the criminalization of AI-generated CSAM.

**4.2. The landscape within the European Union**

**4.2.1. European Union**

The existing legal framework on CSAM offences within the European Union was initially agreed in 2011. According to the article 2(c) iv of the Directive 2011/93, "child pornography definition includes realistic images of a child engaging in sexual conduct or realistic images focusing on the genitalia of a child for primarily sexual purposes". For instance, computer generated material, such as drawings or paintings, which do not portray a real child but give the impression that sexual activity with minors is actually taking place are considered illegal [1] under the current

---

[9] INHOPE, Universal Classification Schema. Retrieved from: https://inhope.org/EN/articles/what-is-the-universal-classification-schema#:~:text=The%20Universal%20Classification%20Schema%20will,contribute%20to%20a%20core%20dataset.

European legislative framework. However, the development of augmented virtual reality settings and practices of misusing AI with the aim of creating "deepfakes" have broadened the definition of child sexual abuse material.

The recast of the Directive 2011/93 [20], especially the proposed amendment to the article 2(3)(d) aims to ensure that the CSAM definition covers the AI developments in a technology-neutral way and addresses the criminalization of AI-generated CSAM. An expansion and clarification of the CSAM definition across the European Member States is proposed, in order for CSAM in deepfakes and AI-generated material to be included [21]. Consequently, the provision of article 2(c)(iv) where the CSAM definition includes "realistic images of a child engaged in sexually explicit conduct or realistic images of the sexual organs of a child, for primarily sexual purposes", is proposed to be reformulated as follows: "realistic images, reproductions or representations of a child engaged in sexually explicit conduct or of the sexual organs of a child, for primarily sexual purposes".

In addition, it is worth mentioning that the recast of the Directive provides also for the criminalization of the instruction manuals on how to sexually abuse children, the "pedophile handbooks", as an important step forward in treating the spread of CSAM within the European Union in a methodical and comprehensive way. Accordingly, "any material, regardless of its form, intended to provide advice, guidance or instructions on how to commit child sexual abuse or sexual exploitation or child solicitation", is encompassed within the CSAM definition.

### 4.2.2. Council of the European Union

While the proposal of the recast of the Directive 2011/93 is under examination by the Working Party on Judicial Cooperation in Criminal Matters (COPEN) [22], it is worth mentioning some points highlighted by the Council [23] of the European Union. The Presidency of the Council of the European Union[10] emphasized on the possibility that a deepfake can refer to a real person, who is a bearer of rights and to which protection can be granted, while he also mentioned that the seriousness of this phenomenon is proved by the fact that AI applications are capable of creating realistic images that are indistinguishable from real images and that they are often trained on the basis of real CSAM.

At the same time, among the Presidency's considerations is found the way that AI-generated CSAM can be addressed from the perspective of criminal law and judicial cooperation. For the Presidency, the combat against child sexual abuse entails the criminalization of new emerging technological developments, something that poses various legal questions from the perspective of Union criminal law legal bases. It is also evident that this kind of content hampers the law enforcement operations to identify whether or not there are real children abused.

### 4.2.3. INHOPE Countries in the EU

As from 2021, INHOPE's approach to the phenomenon was underlying the necessity of legislative revisions in the field: "The development of deepfake technology is an example of the need for legislation to be continually developed as the tools which people use to create and share CSAM change" [24]. For the time being, many countries members of the European Union include "digitally generated depictions of child sexual abuse"[11] in their legal definition of CSAM, meaning this kind of material can be reported and removed from the internet like other forms of CSAM [24].

At the same time, no national legislation in any European country where INHOPE hotlines exist specifically regulates AI-generated CSAM [25]. Nonetheless, various national laws can be interpreted to respond to the illegality of AI CSAM. For instance, AI-generated CSAM can be deemed illegal under the Greek and the German law, since they both criminalize virtual CSAM. On the other hand, there are countries, such as Denmark, that criminalize AI-generated CSAM only in cases it appears to be indistinguishable from real CSAM [25].

---

[10] The presidency of the Council rotates among the EU member states every six months. During this six-month period, the presidency chairs meetings at every level in the Council, helping to ensure the continuity of the EU's work in the Council.
[11] Digitally generated CSAM/realistic images representing a minor engaged in sexually explicit conduct.

## 4.3. INHOPE Countries outside the EU

**United Kingdom**

Under the legislation of the UK, AI CSAM is criminal, actionable under the same laws as real CSAM. The respective legal framework includes the Protection of Children Act 1978 [26], which criminalizes acts related to "indecent photograph or pseudo-photograph of a child", and the Coroners and Justice Act 2009, which criminalizes the possession of "a prohibited image of a child". The latter law refers to non-photographic material, such as cartoons or animations. Proving whether an image is AI-generated is not a requirement for criminal prosecution under the Protection of Children Act. The only requirement is to resemble a photograph and be an indecent image of a child [2]. Specifically, in case AI content does not meet the threshold to be deemed as realistic, images are actioned as non-photographic images of child sexual abuse [25].

**United States of America**

Under the Federal law of the United States of America, virtual CSAM is criminalized, in case the respective content is "virtually indistinguishable" from real children [27]. However, according to NCMEC, it is indispensable that federal and state laws be updated to clarify that AI-generated CSAM constitutes illegal material [28], since there is an ongoing discussion with regard to the extent to which existing laws criminalizing CSAM apply adequately to AI-generated CSAM. It is encouraging though that many states have already proceeded to the revision of CSAM related laws[12].

**Australia**

There is no specific provision regulating AI-generated CSAM in the Australian legal system. However, under the Criminal Code Amendment Bill [29], of 2024, the sharing of non-consensual deepfake sexually explicit material is prohibited. Moreover, according to the Online Safety Industry Standard 2024, internet services and platforms deploying or distributing generative AI models will be obliged to implement measures to eliminate CSAM [30].

**South Korea**

Under the provisions of the South Korean Act that criminalizes [31] CSAM-related conducts, only material including "persons or representations that can be obviously perceived as children or youth" fall under the scope of illegal material. The insertion of the phrase in the wording of the law makes computer-generated images, manipulated material and adults appearing to be children subject to the "Child Sexual Abuse Material" term. Nevertheless, the Korean Supreme Court ruled in 2014 that the CSAM term includes only material where the depicted character "clearly and undoubtedly" appears to be a child and is apprehended as such by the viewer [32]. On October 10th, 2024, the government of South Korea approved the amendment of the Special Act on the Punishment of Sexual Crimes, to include penalties for the crimes of possession and viewing of deepfake exploitation content [33].

**Taiwan**

In early 2023, 4 laws underwent amendments, including the Criminal Code, the Child and Youth Sexual Exploitation Prevention Act, the Sexual Assault Crime Prevention Act and the Crime Victim Protection Act, in order for computer generated CSAM to be criminalized [34]. Thanks to the aforementioned amendments, digitally generated CSAM is deemed illegal in case of a realistic depiction [25].

**Russia**

The preparation of a bill had been in the pipeline, according to which the use of artificial intelligence for the purpose of committing crimes may be included in the list [35] of aggravating circumstances [36]. However, the Russian government did not support the bill considering that the illegal distribution of a person's personal data falls under the article 137 of the Russian Criminal Code and does not require additional regulation [37]. Currently, there is no specific legal provision criminalizing the AI-generated CSAM. Nevertheless, whoever creates or distributes AI-generated CSAM is legally liable under the Russian legislative framework [25].

---

[12] Currently NCMEC is tracking over 38 state laws in 24 different states relating to GAI CSAM. However, there are still legislations that require a real child to be depicted. See:
https://www.proquest.com/docview/3038945924/7D379D783820418APQ/64?sourcetype=Wire%20Feeds.

**Bosnia and Herzegovina**

Under the Criminal Code of Bosnia and Herzegovina [38] "the abuse of a child or juvenile for pornography" is criminalized. However, the Criminal Code of the Srpska Republic criminalizes sexual abuse material which involves a person who "looks like a child" in real or explicitly simulated evident sexual behavior [39] [40]. The Criminal Law of Brčko District [41] is aligned with the aforementioned approach, since "a realistically depicted non-existent child or a person who looks like a child in real or simulated sexually explicit behavior" falls under the CSAM's legal definition. With regard to the category of AI-generated CSAM, there is no relevant provision in any of the above legislative frameworks.

**Brazil**

The Brazilian legal framework does not explicitly criminalize AI-generated CSAM. However, under the Brazilian Child and Adolescent Statute [42] simulating the participation of a child in a pornographic scene by means of montage or modification of a photograph or video constitutes a criminal offence.

**Japan**

We should mention that the Japanese legislation criminalizes pornography depicting existing children, since the law refers to "real people under the age of 18" [43]. Due to the law's wording, the respective Act [44] is only applicable to "real" children [45]. At the same time, under the article 3 of the Act, it is clarified that manga creations are excluded from the "child pornography" definition. That is because "cultural and artistic activities" refer to the Japanese Act for the Promotion of Culture and the Arts, in which manga is acknowledged as significant medium of national tradition [45].

Yet, however, it is imprecise whether virtual pseudo-photographs constitute CSAM. In 2016, indication regarding the legal status was presented in a ruling, concerning creation and sale of computer graphic images of bare youths [46]. The ruling by Tokyo District Court judged that "the composite photographs, pseudo pornography or collage created by using multiple pictures also even if they synthesized a picture of a naked body of a child named B on the face of child A, as long as the child actually exists, the image is a depiction of the appearance of a 'real' child B, and should be interpreted as 'child pornography'" [47]. Despite the fact that the aforementioned ruling is confined to a specific condition, it is implied that virtual content may be interpreted as CSAM. Hence, "any representation" requires interpretation according to the wide definition of the OPSC (Optional Protocol to the Convention on the Rights of the Child on the sale of children, child prostitution and child pornography), which extends to drawings and virtual child pornography [48].

**Turkey**

CSAM crimes are not explicitly criminalized under the Turkish Penal Code. CSAM-related conduct falls nevertheless under the scope of the Code's "obscenity" provision, since under the article 226(3) the exploitation of children in the production of "obscene" audio-visual materials or with the aim of committing crimes, such as selling such material, possessing or sharing is a criminal offence. The fact, however, that the Turkish Penal Code does not define the word "obscene" entails legal uncertainty.

On the one hand, given the broad scope of the "obscenity" term, AI-generated CSAM could be construed as part of the nature of obscene materials, upon legal interpretation. On the other hand, the ambiguity that the term "obscene" creates has been criticized since it sometimes leads to free press restrictions, while it can also hinder children's protection from CSAM crimes [49].

**Thailand**

In terms of the characteristics of child pornographic materials, the Thailand's Criminal Code provision [50] sets the criterion that, if the material in question presents content that can be understood as or depicts sexual acts of a child or with a child in an obscene manner, it is categorized as child pornography [51].

However, the Handbook on the Implementation and Preparation for Law Enforcement in accordance with the Criminal Code Amendment Act, which was produced by the Ministry of Justice of Thailand, makes it clear that the legislators intended to limit the scope of child pornography law to materials depicting real children. Consequently, cartoons, animations and computer-generated images are excluded from the child pornography legal definition [52]. Nonetheless, it is argued that Thailand's current legislation can be interpreted to cover realistic AI-generated CSAM [25], given the broad child pornography definition, which states that "any form" of such material is deemed illegal.

**Philippines**

Philippines CSAM legislation does not include any specific provision on AI-generated CSAM [53]. However, computer-generated content, digitally or manually crafted images or graphics of a person who is made to appear to be a child is illegal under the Philippines legal framework. Additionally, the law criminalizes "image-based sexual abuse" (ISA), which consists of a pattern of behaviors, involving among others the use of AI to create deepfake pornographic videos and the nonconsensual distribution or threat to distribute sexual or nude images of someone [25] [54].

**Colombia**

Under the article 218 of the Colombian Penal Code [55] CSAM-related conducts are criminalized in case the material in question depicts real minors. Therefore, pornographic material showing an adult who simulates being an underage person, cartoons or animation, as well as digitally generated or manipulated images, the production of which does not presuppose any kind of exploitation of a real child, are not explicitly criminalized under the Colombian law. Given the technological advancement, it is highly demanding to proceed to legal reforms in order for the legislative framework to encompass simulated representations of minors and specifically AI-generated CSAM.

**Cambodia**

The CSAM definition provided by the Cambodian law on Suppression of Human Trafficking and Sexual Exploitation [56] does not criminalize virtual child sexual abuse content [57]. The absence of a regulatory framework for the production of virtual CSAM and other virtual CSAM-related conducts seems to pose a severe impediment in the protection against AI-generated material.

**New Zealand**

Most CSAM-related offences are regulated under the New Zealand's Films, Videos, and Publications Classification Act (FVPC Act), as conducts pertaining to "objectionable publications" [58]. Under the FVPC Act's amendment that took place in 2005, CSAM definition [59] extended to "a representation, by any means, of a person who is or appears to be under 18 years of age engaged in real or simulated explicit sexual activities; or a representation of the sexual parts of a person of that kind for primarily sexual purposes".

Objectionable publications range from photographs to disks or electronic computer files, on which information may be stored, recorded, reproduced or shown as one or more images, representation, signs, statements or words [59]. It is worth noting that the New Zealand Law Enforcement operations extend to seizing "disturbingly realistic" AI-generated CSAM [60].

**Ukraine**

The Ukrainian Criminal Code neither mentions AI-generated materials, nor explicitly criminalizes virtual CSAM. Even though the wording of the articles 301 and 301-1 of the Ukrainian Criminal Code [61] covers "child pornography" in a broad sense, digitally generated CSAM in the Ukrainian legal order is not illegal [25].

**Albania**

The Albanian Criminal Code inserts a broad definition of "child pornography" without specifying the different categories of the illegal content. The wording of the law explicitly criminalizes "pornography [62]" involving children, without explicitly addressing content involving real children or fictitious characters. Nevertheless, it is argued that the legality of digitally generated CSAM depends on the realism of the content [25].

**Serbia**

The Serbian legal framework does not criminalize AI-generated CSAM, the article 185 of the Criminal Code though criminalizes not only material depicting real children, but also virtual content created by the use of technology, such as animation or any kind of content appearing to involve children [63].

**Argentina**

Argentine law [64] takes a broad approach to protecting minors, not only prohibiting the use of real children in this type of material, but also criminalizing the production and distribution of content that appears to involve minors, such as in the case of animations, computer-generated images or other types of graphic representations. Specifically, it has been pointed out that the word "representation" includes any simulated image in which a child is depicted in explicit sexual activities or content in which a child's genital parts are being displayed [65]. Even though the wording of the article can possibly cover many different variables of CSAM-related conducts and could likely include any technological means that will be launched in the future, it is argued that terms are left imprecise contrary to the principle of legality [65].

In spite of the absence of specific legislation regulating AI-generated CSAM in Argentina, it is claimed that the use of a person's image without consent to create sexual content, could amount to an illegal intrusion to one's image, which constitutes a highly personal right [25]. In case the person depicted is a minor, the aforementioned act could constitute a crime which could fall under the article 128 of the Argentinian Penal Code [25].

**Moldova**

According to the Article 208 of the Criminal Code of the Republic of Moldova [66], child pornography is defined as any material that explicitly shows minors in sexual activities, including any representation of children involved in explicit, real or simulated sexual activities. However, digitally generated CSAM is not illegal under the Moldovan law [25].

**Iceland**

Under the Icelandic Criminal Code [67] the penalty imposed to offenders for CSAM crimes that involve real children, applies also to CSAM offences where the material concerned depicts adults that impersonate children, cartoons or other virtual images [68]. Concerning the category of digitally generated CSAM, it is treated as illegal under the Icelandic legal order [25], while there are no specific legal provisions covering AI CSAM.

**Nigeria**

Under the Nigerian legal framework, different legal provisions are dedicated to children's protection from abuse and CSAM crimes [69]. At the same time, there are two different criminal codes applicable in the northern and in the southern states. Nevertheless, the notion of CSAM is not explicitly defined under the Nigerian legislation. Concerning AI-generated CSAM, there are no specific laws addressing the phenomenon, however discussions to amend the legislative framework in order to include recent technological advancements have been initiated [25].

**Mexico**

Under the Mexican law drawings and digitally generated CSAM do not constitute illegal material since they do not portray real minors [25]. The legality of this category of content depends on the judge's discretion via supplementary interpretation or case law. Even though there is no specific legislation addressing AI-generated CSAM, one initiative has been launched to introduce ethical regulations for diverse uses of artificial intelligence across the country [70].

This initiative neither addresses fully the phenomenon, nor it specifies whether the AI CSAM constitutes illegal material. Despite that, it constitutes an initial step towards the regulation of the artificial intelligence field.

**South Africa**

Under the South African legislative framework there are no provisions regulating AI-generated CSAM. However, digitally generated CSAM is illegal [25], since the definition of "child pornography" includes "any image, however created, or any description or presentation of a person, real or simulated, who is, or who is depicted or described or presented as being, under the age of 18 years, of an explicit or sexual nature" [71].

Table 2: Summary on legal developments in the AI field and the criminalization of AI-CSAM crimes in INHOPE countries

| Country / Entity | AI-specific Law or Policy / Specific Criminal Provision on AI-generated CSAM |
|---|---|
| Cambodia | There are no national policy initiatives. There only exists the ASEAN Guide on AI Governance and Ethics (2024) which constitutes a practical guide for organizations in the region that wish to develop traditional AI technologies in commercial and non-military or dual-use applications [72]. |
| New Zealand | There is no AI-specific national regulation. However, since 2020, the Algorithm Charter for Aotearoa New Zealand has been launched. The Charter is a framework aiming at laying out principles for responsible and trustworthy use of algorithms by government agencies of New Zealand [73]. |
| Colombia | There are two Declarations signed in Colombia concerning AI regulation. The first one is Montevideo Declaration which was signed by the Ministers and High-Level Authorities representing the countries gathered at the Second Ministerial and High-Level Authorities Summit on the Ethics of Artificial Intelligence in Latin America and the Caribbean. The Declaration demonstrates the region's commitment to deploy AI systems based on the respect of fundamental rights, while promoting innovation [74].<br><br>The second one is the Santiago Declaration to promote Ethical Artificial Intelligence in Latin America and the Caribbean [75]. The Santiago Declaration incorporates UNESCO's Recommendation on the Ethics of Artificial Intelligence, while it includes the principles that should underline public policy on AI. |
| Serbia | X |
| Albania | X |
| South Africa | In South Africa there is no AI legislation. However, the National AI Policy Framework was established during 2024, which outlines the main strategic pillars for AI in the country [76].<br><br>In parallel, South Africa's AI Plan was released in 2023 [77]) **[DRAFT]** as a draft discussion document and includes the country's strategy for AI on many different sectors. |
| Mexico | Law for the Ethical Regulation of AI for the Mexican United States **[DRAFT]** [78]. |
| Nigeria | National AI Strategy **[DRAFT]** [79]. |
| Iceland | Iceland's national AI policy has been released since 2021. It includes key objectives for building an ethical basis for the development and use of AI [80]. |
| Moldova | Moldova's legal order does not include any AI legislative framework for the time being.<br><br>However, the country holds an Association Agreement with the EU, according to which the country often harmonizes its legal framework with the EU acquis [81]. There is therefore a high likelihood that Moldova will follow the European legislative framework in the future. |
| **Ukraine** | Within the Ukrainian jurisdiction, there are no AI-specific regulations.<br>However, in the soft law[13] context, the national Ministry of Digital Transformation has issued a Whitepaper on AI [82].<br>There also exists the AI Roadmap [83] which comprises of the Guidelines of the Responsible Use of AI in the News Media (2024) [84] and the development of Ukraine's AI law. |
| **Philippines** | House Bill No. 10944 (Artificial Intelligence Act), 2024 **[DRAFT]**: Introduces a regulatory framework for the deployment and use of artificial intelligence, establishes the AI Board and outlines the roles of various government agencies. The Draft Bill also criminalizes certain prohibited conducts [85].<br>House Bill No. 9425 (An Act Defining Deepfake and the Imposition of Penalties Therefor for Acts Constituting Harmful Deepfakes), 2023 **[DRAFT]**: The Draft Bill criminalizes the use of the deepfake technique in committing certain crimes, like cyber fraud and gender-based sexual harassment [86].<br>House Bill No. 10567 (Deepfake Accountability and Transparency Act), 2024 **[DRAFT]**: Under this Draft House Bill, any person producing, distributing, or creating a deepfake with the intent to distribute it over the internet, shall guarantee that the deepfake includes a disclosure in the form of audio, visual, or audiovisual content [87].<br>House Bill No. 7396 (Act promoting the Development and Regulation of Artificial Intelligence in the Philippines), 2023 **[DRAFT]**: This House Bill proposes the establishment of the Artificial Intelligence Development Authority ("AIDA") [88].<br>--------<br>Under the Anti-Online Sexual Abuse or Exploitation of Children and Anti-Child Sexual Abuse or Exploitation Material Act – Republic Act No. 11930, image-based sexual abuse, including the use of AI to create deepfake pornographic videos, is criminalized [54]. |
| **Thailand** | AI Ethics Guidelines (2021) [89].<br>Royal Decree on AI System Service Business (2022) **[DRAFT]**: Proposal for the regulation of AI systems on a risk-based approach [90]. |
| **Argentina** | AI National Plan (2019): National Strategy for the period 2020-2030 including policies applicable to AI development [91].<br>Cartagena Declaration on AI Governance (2024): AI Declaration was adopted by 17 Latin American States [92].<br>Ibero-American Charter of Artificial Intelligence in Public Administration (2023) [93]. |
| **UK** | UK does not intend to implement horizontal AI Regulation [94]. On the contrary, the UK government's White Paper and its written response to the feedback it received ("The Response") indicate that regulators are left to interpret and implement various AI ethics principles in the respective domain [95].<br>Public Authority Algorithmic and Automated Decision-Making Systems Bill (2024) [96].<br>Introduction to AI assurance (2024): Guidance [97].<br>Generative AI Framework for HMG (2024) [98]: guidance for civil servants and workers in government organizations, related to the secure use of generative AI.<br>National Whitepaper on AI Regulation (2023) [99]: published by the UK Government including principles-based regulatory framework.<br>--------<br>AI CSAM is actionable under the laws of real CSAM:<br>Protection of Children Act 1978 [95] Coroners and Justice Act 2009 [100]. |
| **Australia** | In Australia there is no AI-specific Regulation yet. Currently, there exist the National Framework for the assurance of artificial intelligence in government [101] and the Australia's AI Ethics Principles [102]. |

---

[13] The term "soft law" refers to non-legally binding agreements, principles, guidelines, or declarations that, while not legally enforceable like treaties and statutes, still influence decision-making processes within various legal frameworks.

| | |
|---|---|
| | The Criminal Code Amendment (Deepfake Sexual Material) Bill [29], of 2024, amends the Criminal Code Act 1995, while introducing new criminal offences. Specifically, it bans the sharing of non-consensual deepfake sexually explicit material.<br>According to the Online Safety Industry Standard 2024, internet services and platforms deploying or distributing generative AI models will be obliged to implement measures to eliminate child sexual abuse material [30]. |
| **USA** | Artificial Intelligence Civil Rights Act (2024) **[DRAFT]**: This bill seeks to regulate AI algorithms used in decisions in various sectors including employment, banking, healthcare, criminal justice and public services [103].<br>Content Origin Protection and Integrity from Edited and Deepfaked Media Act (COPIED Act) (2024) **[DRAFT]**: introduced to ensure transparency and content provenance [104].<br>Child Exploitation and Artificial Intelligence Expert Commission Act (2024) **[DRAFT]**: Legal framework aiming at assisting law enforcement in detecting and prosecuting AI-generated crimes against children [105].<br>Over 38 State Laws have been adopted in 24 different States concerning Generative AI CSAM [106]. |
| **Bosnia Herzegovina** | X |
| **South Korea** | Basic Act on the Development of Artificial Intelligence and the Establishment of Trust (2024) [107].<br>The creation of sexually explicit deepfakes with the intention to distribute them is deemed illegal under the Sexual Violence Prevention and Victims Protection Act (2024) [108] [109].<br>On October 10$^{th}$, 2024, the government of South Korea approved the amendment of the Special Act on the Punishment of Sexual Crimes, in order to include penalties for the crimes of possession and viewing of deepfake exploitation content [33]. |
| **Taiwan** | AI Basic Act (2023) **[DRAFT]**: Proposal for an AI Regulation [110]. |
| **Turkey** | Artificial Intelligence Law (2/2234) **[DRAFT]**.<br>National Artificial Intelligence Strategy (2021-2025) 2021 [111]. |
| **Russia** | Bill establishing Digital Innovation and AI in Experimental Legal Regimes (Bill No. 512628-8) (2023) [112].<br>Russian Law No. 258-FZ (2021) on the establishment of regulatory sandboxes across Russia [113].<br>Code of Ethics of AI (2021) [114].<br>Federal Law No. 34-FZ (2019) [115]: Known as the Digital Rights Law, aims at the establishment of a legal framework of digital economy in Russia, taking into consideration digital technologies, big data challenges and the AI landscape.<br>Currently, there is no specific legal provision criminalizing AI-generated CSAM. Whoever creates or distributes AI-generated CSAM, though, is legally liable under the Russian legislative framework [25]. |
| **Japan** | AI Guidelines (2023) [116].<br>National Strategy in the New Era of AI (2023) [117].<br>Governance Guidelines for the Practice of AI Principles (2022) [118].<br>Social Principles of Human-Centric AI (2019) [119]. |
| **Brazil** | Artificial Intelligence (AI) Plan (PBIA) 2024-2028 [120].<br>Bill 2807/2024 [121] **[DRAFT]**: aims at banning the insertion of children's images in AI systems, without their parents' or legal guardians' explicit consent.<br>Cartagena Declaration on AI Governance (2024): AI Declaration was adopted by 17 Latin American States [92]. |

| **European Union (EU)** | On 1 August 2024, the European Artificial Intelligence Act entered into force [122]. The AI Act aims to guarantee responsible AI deployment within the EU, by addressing potential risks to citizens' fundamental rights. It also provides developers and deployers with clear obligations concerning specific AI uses, while it also regulates general purpose AI models. <br><br> Directive on a liability regime for AI **[DRAFT]** [123]. This proposed directive intends to guarantee that people harmed by AI systems receive the same level of protection as people harmed by other technologies in the European Union. <br><br> Ethics guidelines for trustworthy AI [124]. |
|---|---|
| | The Recast Directive [125] **[DRAFT]** introduces higher penalties for sexual crimes against children, broadens the definition of child sexual abuse material (CSAM) to include 'pedophile manuals' and AI-generated content, while it also creates a legal basis for INHOPE hotlines of the EU to process CSAM reports. |

### 4.4. Other Countries

**North America**

Sexually explicit content of a minor is criminalized by the Canadian law, which explicitly criminalizes material that has been altered by AI technology [126]. Some Canadian provinces have even adopted additional legislations that allow victims to bring a civil action in court against the offender [127]. It is worth noting that Canada's national Tipline for reporting the online sexual exploitation of children is a former member of the INHOPE network [128].

**Latin America**

According to ICMEC's Guidelines for the adoption of national legislation in Latin America [129], only eight countries in Latin America criminalize simulated (virtual) child sexual abuse material: Brazil [43], Costa Rica [130], Dominican Republic [131], Guatemala [132], Mexico [133], Nicaragua [134], Panama [135], Uruguay [136]. However, none of these countries include provisions that expressly criminalize AI-generated CSAM.

**African Countries**

The African Union Convention on Cyber Security and Personal Data Protection obliges member states to establish regulatory measures to control cybercrime [137]. Under the Convention's definitions, various child pornography offences are included, while the "child pornography" term includes computer generated CSAM [138]. Only a few member states, including Ghana, Guinea, Namibia, Mauritius and Senegal, have ratified the African Union Convention on Cybersecurity and the Protection of Personal Data, known as the "Malabo Convention" [138]. However, it seems that countries do not always meet the requirements set by these international treaties.

Some national legislations though have vague child pornography definitions that seem to include virtual material as well. One such example is the Senegalese [139] child pornography laws [140]. On the other side, some legislations integrate apparent reference to computer-generated images. This is the case of the South African legislation [141]. However, there is currently no legislative provision explicitly relating to AI-generated CSAM.

**Asian Countries**

None of the Asian countries have adopted legislative provisions including explicitly AI-generated CSAM in the legal definition of CSAM. However, some legislations encompass vague CSAM definitions, which upon legal interpretation could probably include AI-generated CSAM. For instance, Israeli Penal Code criminalizes the obscene publications [142], but does not define "obscene" nor does it elucidate on meaning of the phrase "representation or a drawing of a minor" [143]. As a result, the degree of discretion granted to the Israeli Judiciary makes it difficult to ascertain whether Israel is fully complying with the OPSC. "Pornographic materials and obscenity against children" are criminalized under the Vietnamese legislation [144], while obscene publications exploiting the nudity of children also constitute a crime under the law of Singapore[14].

On the other side, there exist more comprehensive CSAM definitions, such as, for instance, the Indian CSAM legislation. Under the POSCO Act, "child pornography" refers to any visual depiction of sexually explicit conduct involving a child, any photograph, video or digital and computer-generated image that is indistinguishable from an actual child, and any image that is modified or adapted, but appears to depict a child [145]. It is claimed that the aforementioned comprehensive CSAM definition may serve as a paradigm to address the AI-generated CSAM phenomenon [146]. Accordingly, the Indonesian Law on Pornography defines CSAM as "all kinds of pornography that involve a child or include an adult who acts like a child [147]". At the same time, the Penal Code criminalizes conduct related to writings, portraits or objects offensive to decency [148]. The vague wording of this provision leaves the possibility of invoking it for CSAM-related crimes when it comes to judicial interpretation [149]. Furthermore, the Penal Code criminalizes the public exhibition or display of writings or portraits of children under 17 years of age that "arouse or stimulate the sensuality" [148]. Regarding legislative developments in China, it should be noted that steps have been taken towards the adoption of a legal framework regulating the deepfake technology. The respective draft Regulation's [150] key points include the responsibility of online content providers to use labels for ai-generated content, such as metadata tags or digital watermarks, while no one should maliciously delete or distort these required labels that aim at identifying ai-generated content.

**4.5. Other Legislation that affects the regulation of AI-generated CSAM**

Along with the analysis of each country's legislative vacuum and criminal provisions, it is worth mentioning that the advent of another legislative framework may affect and enable the production of sexually abusive content. The European AI Act [122] establishes a common regulatory and legal framework for AI within the EU. Among its various legislative provisions, the AI classification system is established, which introduces the "unacceptable risk" category, encompassing systems considered to be dangerous or harmful to fundamental rights and safety.

Despite the fact that AI Act prohibits some harmful practices, it is argued that [151] it fails to regulate the growing problem of deepfake extortion or AI-generated CSAM, which clearly constitute fundamental rights' violations. The AI Act does not include the AI systems used to generate deepfakes in the "high-risk" category. It assigns them instead to a lower-risk category and only imposes on them transparency obligations, such as content disclosure used for training purposes of the general-purpose AI models, labeling of content generated as a deepfake and detection measures that would ensure that deepfake content would be identified and flagged. Thus, it is supported that AI-generated CSAM should be addressed with stricter regulative response, within the AIA, possibly with an elevated level of categorization.

**5. Conclusions**

In conclusion, it is obvious that the absence of a clear legal framework worldwide combined with the increasing realism of AI-generated CSAM presents significant challenges for CSAM offences prosecution and content moderation. The process of identifying CSAM online and its removal from the internet may become extremely slow, while the existence of improved AI models, together with the growing communities that share AI content, render it even more difficult to effectively address the phenomenon. Notwithstanding the fact that a homogeneous framework will not be established in the foreseeable future, we should keep in mind that even if the respective criminal offence does not include a real minor, the aim should be to punish behaviors that form part of a subculture that favors sexual abuse of minors.

The lack of international legal standardization of sanctioning when it comes to material such as "pseudo photographs" and "virtual CSAM" combined with the easiness of creating deepfakes in this legislative vacuum [152], also hamper the investigations of potentially real abuse victims. Although some AI companies have already adopted measures to prevent this phenomenon, open-source platforms have been much slower to take such safeguards into account, while their systems facilitate the use of real children's images as source material. Consequently, the impact of generative AI will depend on how it will be regulated, given that AI-generated CSAM frequently depicts fictitious children, which constitutes a fact that blurs the lines of legality. In terms of prevention though, policy makers should always bear in mind that the AI-generated content entails revictimization of known child sexual abuse victims, while it definitely causes distress to children whose images are being used in a sexualized manner. Certainly, however, we are reaching the conclusion that serious violations of children's human rights are taking place, which are protected by the International Convention on the Rights of the Child, as well as by the European Convention on Human Rights and Fundamental Freedoms.

---

[14] Criminalized under the "Undesirable Publications Act".

With illegal activity continuing unabated on the dark web and the clear establishment and spread of more sophisticated ways to exploit existing artificial intelligence models, it is certain that a holistic approach must be developed in the fight against various forms of child sexual abuse and exploitation material online. Beyond regulatory efforts to prevent and prosecute such crimes against children, the adoption of strict standards in the areas of safety and the prevention of the exploitation of open-source software for malicious purposes is deemed necessary, alongside the evaluation and continuous reassessment of the systemic risks of platforms, at a time when the capabilities of artificial intelligence are beyond human perception.

**Acknowledgments**

This work was supported by the SI4KIDS.GR (GA No 101146190) 2024-2026, EU project and the PreventCSAatEU (Towards a Coordinated and Cooperative Effort for the Prevention of Child Sexual Abuse at a European level) project (MIS 6002910) 2024-2026.